\documentclass[journal=ancac3,manuscript=article]{achemso}

\usepackage{morefloats}
\usepackage{color}
\usepackage{epsfig,graphicx,amsfonts,amsbsy}
\usepackage{amsmath,amsfonts,amsthm,amssymb}
\usepackage{appendix}
\usepackage{bbm}
\usepackage{makeidx}
\usepackage{url}
\usepackage{verbatim}
\usepackage{mathrsfs} 
\usepackage{morefloats}
\usepackage{appendix}
\usepackage{bbm}
\usepackage{makeidx}
\usepackage{url}
\usepackage{verbatim}

\usepackage[bookmarksnumbered,pdfpagelabels=true,plainpages=false,colorlinks=true,linkcolor=black,citecolor=black,urlcolor=black]{hyperref}

\usepackage{array}
\usepackage{booktabs}
\usepackage{multirow}
\usepackage{bbm}
\usepackage{tabularx}
\usepackage{cancel,soul}
\usepackage{nicefrac, xfrac}
\usepackage{ulem}
\usepackage{bbold}

\usepackage{xcolor}
\usepackage{nicefrac, xfrac}
\usepackage[version=3]{mhchem} 

\def \usach {Departamento de F\'isica, Universidad de Santiago de Chile, Santiago, 9170124, Chile.}
\def \cedenna {Centro  de Nanociencia y Nanotecnología CEDENNA, Avda. Ecuador 3493, Santiago, 9170124, Chile.}
\def \fcfm {Departamento de F\'isica, FCFM, Universidad de Chile, Santiago, 8370448, Chile.}

\def \AUtonUnv {Grupo de Investigación en F\'isica Aplicada (GIFA), Facultad de Ingenier\'ia, Universidad Aut\'onoma de Chile, Av. Pedro de Valdivia 425, Providencia, Santiago, 7500912, Chile.}

\def \udesarrollo {Facultad de Ingen\'ieria, Universidad del Desarrollo, Av. Plaza 680, Las Condes, Chile  }

\author{Jose Toledo-Marin}
\affiliation{\usach}
\alsoaffiliation[\cedenna]{\cedenna}

\author{Mario Castro}
\affiliation{\fcfm}

\author{David Galvez-Poblete}
\affiliation{\usach}
\alsoaffiliation[\cedenna]{\cedenna}

\author{Bruno Grossi}
\affiliation{\udesarrollo}

\author{Sebastián Castillo-Sepúlveda}
\affiliation{\AUtonUnv}

\author{Alvaro S. Nunez}
\affiliation{\fcfm}

\author{Sebastian Allende}
\affiliation{\usach}
\alsoaffiliation[\cedenna]{\cedenna}
\email{sebastian.allende@usach.cl}

\title
  {Magnetic Worms: Oscillatory Bimeron Pairing And Collective Transport In Patterned Stripes}


\begin{document}


\begin{abstract}
Magnetic bimerons in a domain wall provide a practical route for current driven
transport in patterned magnetic stripes. However, coupling between bimerons and
pinning by defects complicate reliable motion. Here we show that a periodic
array of edge defects both stabilizes transport of multiple bimerons and gives
rise to a distinctive collective state, the magnetic worm. A single
bimeron travels at constant speed; defects lower this speed while preserving an
approximately linear relation between velocity $v$ and current density $J$.
With two bimerons, the center of mass advances nearly uniformly while their
separation exhibits a bounded oscillation whose frequency increases and amplitude decreases with current. For larger trains, these oscillations lose synchrony, producing segmented, worm like motion. The center of mass speed grows with current but decreases as the number of bimerons increases. Notably, eight bimerons cannot be sustained in a smooth stripe but can be stabilized by the periodic defects. 
\end{abstract}

\section{Introduction}

In recent years, the field of spintronics has attracted significant attention due to the potential to exploit magnetic textures and spin-polarized currents to transport, manipulate and store information in spintronic devices, with the prospect of enhancing current  \cite{Hirohata2020,Yu2021}. Within this context, racetrack devices stand out as particularly promising\cite{Parkin2008,Parkin2015,Gu2022}. The core concept of these systems lies in the ability to transport magnetic solitons, viewed as information carriers, along a strip by means of spin-polarized currents or applied magnetic fields, enabling controllable and efficient operation. Among the magnetic solitons considered for this purpose are domain walls\cite{Hayashi2008,Han2009,Emori2013,Je2013}, vortices\cite{Geng2017,Antos2008,Choe2004}, magnetic skyrmions\cite{Fert2013,Woo2016,Romming2013,Muhlbauer2009,Tomasello2014,Zhang2015}, bimerons\cite{Gbel2019,Castro2023,Zhang2020,Shen2020,Li2020,SilvaJunior2024}, and others. These magnetic structures stand as promising candidates due to their topological stability, nanoscale size, and ability to be manipulated by electric currents or magnetic fields\cite{Nagaosa2013}.

Despite their great potential for applications, magnetic textures in realistic materials face significant challenges due to inhomogeneities, impurities, and structural imperfections, which act as pinning centers and strongly influence soliton dynamics. In domain walls, such defects can locally alter the internal structure, modify the effective width, and generate spatially varying pinning potentials, leading to complex motion, including creep, intermittent depinning, and mobility changes, that depends sensitively on defect density, distribution, and interaction strength\cite{Thomas2006}. Moreover, domain walls require high current densities to reach efficient velocities and are subject to the Walker breakdown\cite{Schryer1974,Mougin2007}. Vortices and skyrmions, while needing lower current densities, experience the Skyrmion Hall effect\cite{Iwasaki2013,Litzius2016,Jiang2016} due to their topology; the resulting transverse force deflects their trajectory toward the edges, where annihilation occurs. Some textures, such as bimerons, can avoid the edge problem induced by the Skyrmion Hall effect\cite{Castro2025}, and the use of a longitudinal domain wall has been shown to protect skyrmions from edge-related annihilation\cite{Xing2020,Song2020,Yang2022,Nie2025}. In systems hosting more than one magnetic texture, whether skyrmions, bimerons, or domain walls, the interplay between the driving force, damping, and defect-induced pinning generally leads, after a brief transient, to a velocity-matched state with no sustained relative oscillations. In this conventional scenario, defects set the average mobility and introduce local distortions, but do not sustain long-lived oscillatory motion between textures, resulting instead in rigid, collective translation along the racetrack\cite{Zhang20152}.

In this work, we propose a racetrack-like system consisting of a magnetic bimeron embedded within a longitudinal domain wall\cite{Saji2025,Chen2025, https://doi.org/10.48550/arxiv.2505.00959}, demonstrating its ability to be protected against the edge irregularities. We find that, when bimerons are confined within domain walls in a racetrack containing distributed defects, the interplay between defect pinning, inter-soliton interactions, and the internal degrees of freedom of the wall can sustain long-lived relative oscillations. This effect becomes even more striking when two domain walls coexist, as they can oscillate with respect to one another while both remain in net forward motion—an interaction mode not reported previously. Strikingly, when multiple bimerons populate a single domain wall, their oscillations become desynchronized and irregular, giving rise to a collective, segmented motion reminiscent of the crawling gait of a worm. We refer to this phenomenon as the magnetic worm effect. Beyond its conceptual novelty, this dynamic state may have practical implications for the design and operation of racetrack memories and logic devices. The observed interplay between soliton–defect pinning, inter-soliton interactions, and domain wall internal modes highlights new possibilities for engineering controlled oscillatory states in spintronic systems, and calls for a deeper theoretical and experimental investigation into their stability, tunability, and potential device functionalities.

\section{Model}

\begin{figure}[bht!]
    \centering
    \includegraphics[width=0.9\linewidth]{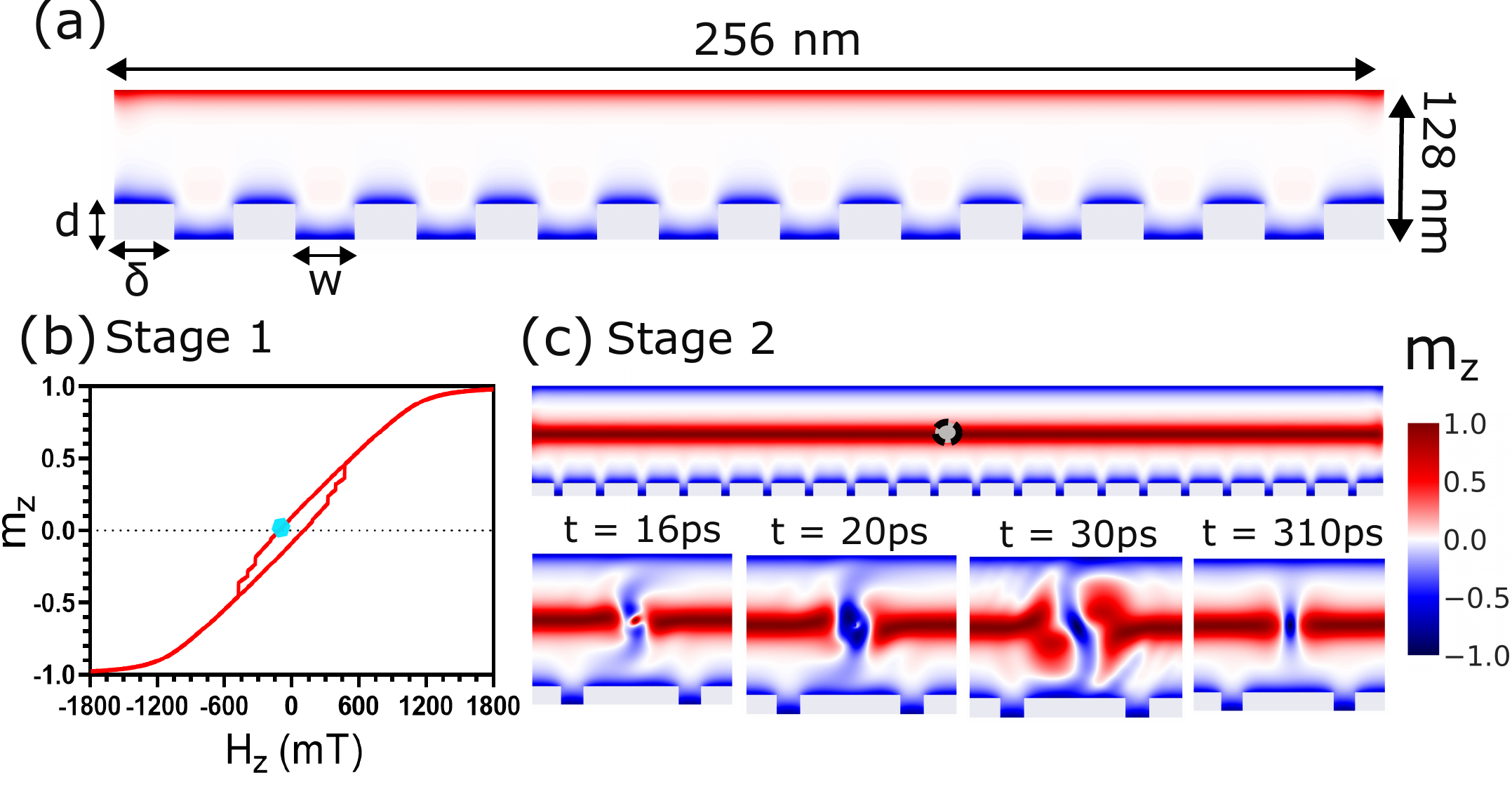}
    \caption{(a) Schematic representation of the patterned nanostripe geometry. The geometry has a length of 2048 nm and a height of 128 nm. Periodic edge defects are introduced along the lower boundary of the nanostripe, characterized by their depth $d$, width $w$, and the separation between notches $\delta$. (b) Out-of-plane hysteresis loop showing the magnetization reversal process. The cyan dot indicates the field value at which the domain wall configuration shown in panel (c) is extracted. (c) The top panel displays the magnetization at $H_z = 0$ llustrating the formation of a longitudinal domain wall at the center of the nanostripe. The bottom snapshots show the time evolution of the magnetization after the injection of a localized current pulse (applied within the circular region marked by the dotted outline), leading to the formation of a bimeron domain wall.}
    \label{fig1}
\end{figure}

We consider a ferromagnetic thin nanostripe of length 2048 nm, height 128 nm, and thickness 1 nm deposited on a heavy-metal underlayer. The system stabilizes a longitudinal domain wall bimeron configuration, aligned along the length of the nanostripe. To investigate the influence of edge inhomogeneities, the lower boundary of the nanostripe is patterned with a periodic sequence of notches (edge defects), characterized by their depth $d$, width $w$, and $\delta$ defined as the center-to-center distance between neighboring notches, as illustrated in \ref{fig1}.(a). To characterize this system, we perform micromagnetic simulations using the open-source micromagnetic framework MuMax$^3$ \cite{Vansteenkiste2014}. The time evolution of the normalized magnetization $\vec{m} = \vec{M}/M_s$, where $M_s$ is the saturation magnetization, is computed by solving the Landau–Lifshitz–Gilbert (LLG) equation\cite{Gilbert2004} including the spin–orbit torque (SOT) term, which is relevant for ferromagnet/heavy-metal bilayer systems\cite{Bttner2017,Liu2012}.

\begin{equation}\label{eq:1}
    \frac{\partial \vec{m}}{\partial t}=-\gamma \vec{m} \times \vec{\rm H}_{\mathrm{eff}}+\alpha \vec{m}\times  \frac{\partial \vec{m}}{\partial t}- J \tau_{DL}\vec{m}\times (\vec{m}\times \vec{m}_p) -  J \tau_{FL}   (\vec{m}\times \vec{m}_p),
\end{equation}

where $\gamma$ is the gyromagnetic ratio, $\alpha$ is the Gilbert damping constant, $\vec{H}_{\text{eff}}$ is the effective fields which includes contributions from  the exchange, uniaxial anisotropy, magnetostatic, and interfacial dzyaloshinskii moriya interaction. The last two terms in the equation account for the damping-like and field-like components of the spin–orbit torque (SOT), respectively, which arise from the interaction between the ferromagnet and the heavy-metal underlayer via the spin Hall effect\cite{Joos2023}.  The damping-like torque coefficient is given by
 $\tau_{DL}= \dfrac{  \gamma \hbar\theta_{SH}}{2  M_s e h} $  while the field-like component is $\tau_{FL} = \xi \tau_{DL}$, where $\hbar$ is the Planck’s constant reduced, $|e|$ is the electron charge, and $\xi$ is the ratio of the magnitudes of the field-like and damping-like terms. Additionally, J is the magnitude of the electric current density and $\vec{m}_p = (0,-1,0)$. For the simulations, we use material parameters corresponding to a Co/Pt bilayer\cite{Saha2019}: saturation magnetization $M_s = 5.8\times 10^5$ A/m, exchange stiffness $A = 1.5 \times 10^{-11}$ J/m, uniaxial anisotropy constant $K_u = 1.4 \times 10^5 $ J/m$^3$ with easy axis along y-direction, interfacial DMI constant $D = 2.9 $ mJ/m$^2$, and spin Hall angle $\theta_{\mathrm{SH}} = 0.19$. For simplicity, the field-like torque is neglected in all simulations ($\xi = 0$), as the interfacial Rashba effect is expected to be weak compared to the spin Hall effect in ferromagnet/heavy-metal bilayer systems \cite{Chen2025}.\\

In ferromagnetic nanostripes, transverse domain walls of Bloch or Néel type can be experimentally nucleated through a variety of mechanisms, including the application of external magnetic fields, spin-polarized current pulses, and geometrical constrictions, among others \cite{Yamaguchi2004, Bhatia2021, Alejos2017}. These methods have been extensively used to initialize controlled domain wall configurations in magnetic nanostructures.  Inspired by these experimental techniques, we employ a two-stage process to nucleate a bimeron domain wall in our system. In the first stage, a  magnetic field is applied along the z-direction and gradually reduced, as shown in Fig.\ref{fig1}.(b). The nanostripe was built considering $d=15$ nm, $w=20$ nm and $\delta = 80$ nm. When $H_z = 0$, a longitudinal domain wall is stabilized at the center of the nanostripe, see top panel of Fig.\ref{fig1} (c). In the second stage, a perpendicular current density of  $J = -5
\times 10^{12}$ (A/m$^2)$ $\sin(2\pi t/T)$ with $T = 0.1$ ns, is applied within a circular region of 30 nm of diameters (indicated by the black dotted circle) during 20 ps . This localized current reverses the magnetization in that area from $\hat{z}$ to $-\hat{z}$ leading the formation of the bimeron domain wall. Similar methods have been employed to nucleate skyrmions and bimerons in other magnetic systems \cite{Belrhazi2022, Castro2025}. In our simulations, for simplicity, the bimeron domain wall is nucleated by defining a small circular zone of 20 nm located along the domain wall with the magnetization set to $-\hat{z}$. After relax, it is displaced toward the lower edge of the nanostripe by applying a static magnetic field of -10 mT $\hat{y}$.

\section{Results and Discussions}

\begin{figure}[ht!]
    \centering
    \includegraphics[width=0.7\linewidth]{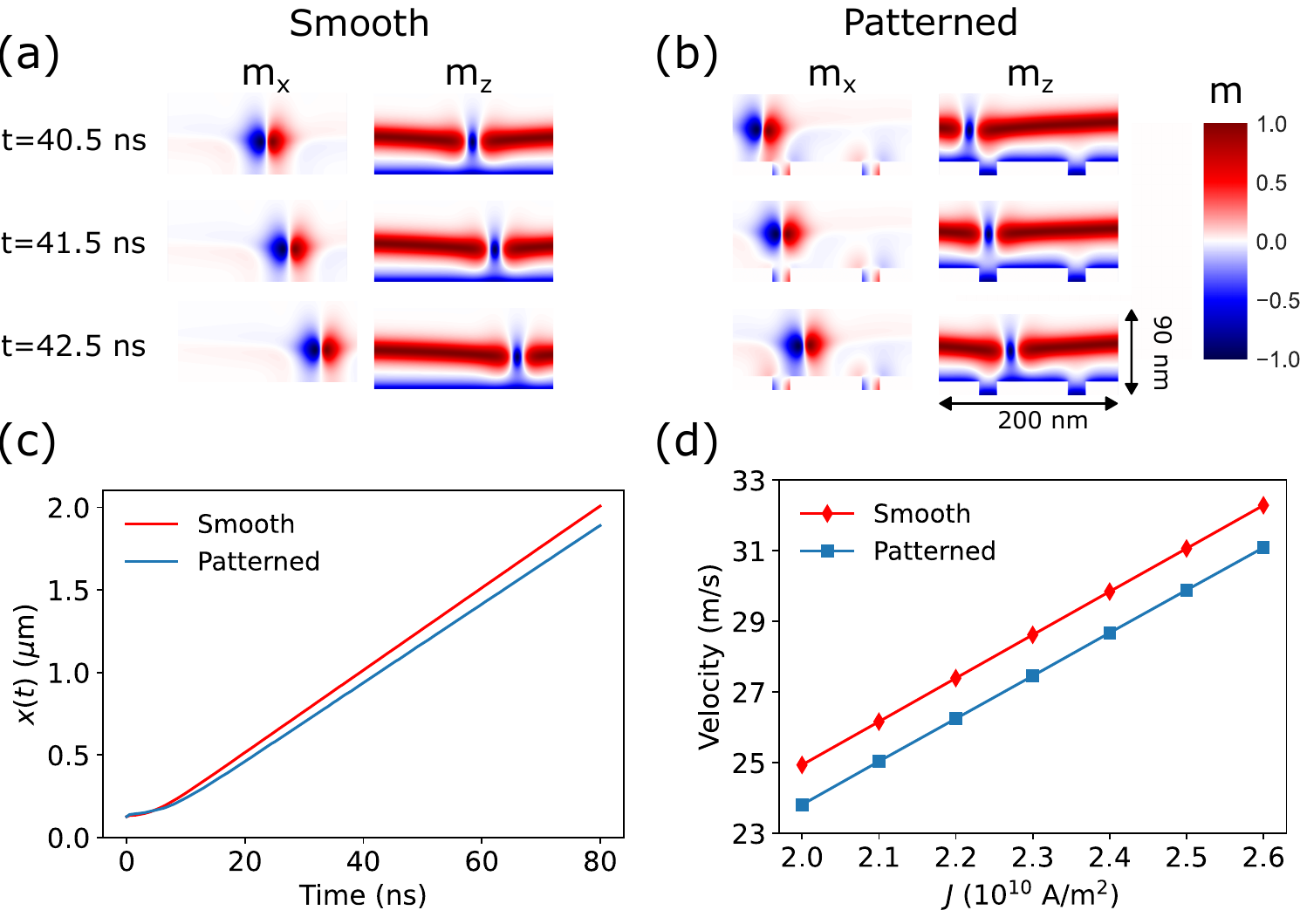}
    \caption{Current-driven dynamics of a bimeron domain wall in smooth and patterned nanostripes. (a) and (b) show the magnetic profile of an isolated bimeron domain wall at different time instants, for smooth and patterned nanostripes, respectively. The magnetic structure remains preserved during the motion in both cases. Each frame covers an area of 200 $\times$ 90 nm$^{2}$.  (c) Position of the bimeron domain wall center as a function of time  for $J = 2\times 10^{10}$ A/m$^2$. (d) Velocity of the bimeron domain wall as a function of the electric current density, comparing smooth and patterned geometries. }
    \label{fig2}
\end{figure}

To understand the construction and behavior of the magnetic worm, we first study how a bimeron behaves within a domain wall. Figure \ref{fig2}.(a) illustrates a bimeron moving along a nanotrack with no defects; after a short transient, it reaches a constant velocity under an applied current density. A key feature is that the bimeron carries opposite magnetic charges along the $y$ direction, which becomes important when a periodic defect is introduced (Patterned).  Figure \ref{fig2}.(b) shows that alternating positive and negative charges form at the notch corners, following the defect pattern in the $y$ component. These charges interact with those of the bimeron and reduce its speed. Figures \ref{fig2}.(c)–(d) quantify this: the displacement $x(t)$ remains linear after the transient, but the slope and hence the velocity is lower in the patterned case. For example, at $J=2.0 \times 10^{10} $ A/m$^2$ the velocity drops from $v=24.94$ m/s (smooth) to $v=23.81$ m/s (patterned), a 4.53\% reduction. In both cases the velocity scales approximately linearly with the applied spin-polarized current density.

\begin{figure}[ht!]
    \centering
    \includegraphics[width=0.7\linewidth]{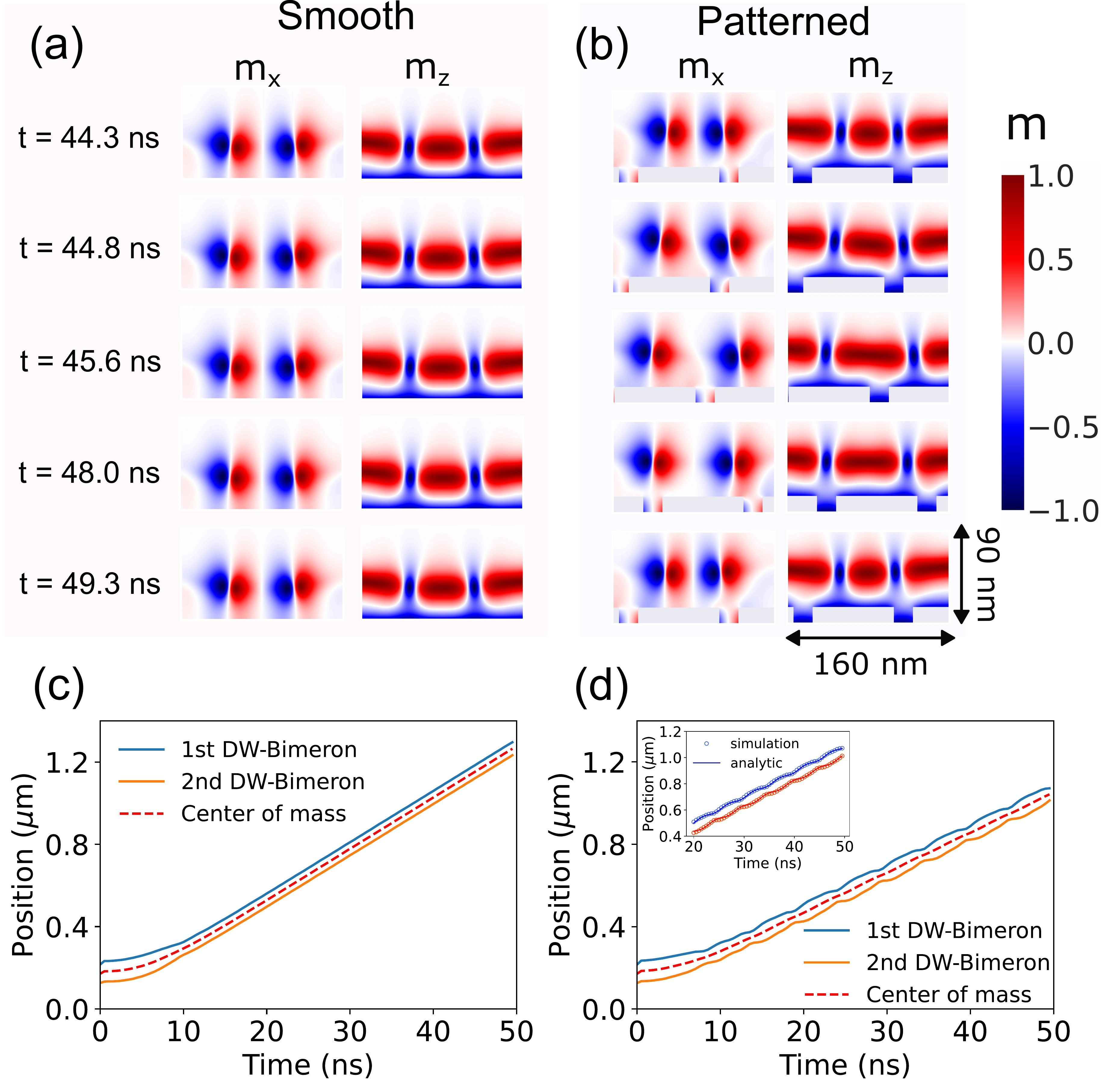}
    \caption{(a) and (b) Magnetic profiles of two domain wall bimerons at different times, stabilized in a smooth nanostripe (left panel) and in a patterned nanostripe (right panel). Each frame covers an area of 160 $\times$ 90 nm$^{2}$. In the smooth case, the bimerons propagate while maintaining an approximately constant separation. In contrast, in the patterned case, their separation exhibits oscillations, alternately approaching and receding, due to the interaction with edge defects. For further clarity about the bimeron domain wall motion see the supplementary video 1 (smooth case) and 2 (patterned case). (c) and (d) shows the positions as a function of time of the center of each bimeron domain wall. The dashed red line represent the position of the center of mass. The inset compare the trajectory obtained from micromagnetic simulations and the simplified model \ref{eq:solutions}. }
    \label{fig3}
\end{figure}

Now we introduce two bimerons into the domain wall. One would expect that, after a short transient, the pair settles at a fixed separation and then translates at constant velocity under a spin-polarized current. This is exactly what we observe on a defect-free track, see figures \ref{fig3}.(a) and \ref{fig3}.(c). These figures show that the distance between the two bimerons remains constant, $\Delta x \simeq 63\,\mathrm{nm}$, for $J = 2.0\times10^{10}\,\mathrm{A/m^{2}}$, while both textures move with the same velocity $v \simeq 24.94\,\mathrm{m/s}$; the dashed curve gives the mass center (cm), which advances linearly in time. Similarly, we expect comparable behavior in the presence of defects, consistent with the result shown in Fig.~\ref{fig2}. However, when a periodic defect is patterned along the track, intuition breaks down. As seen in \ref{fig3}.(b), the pair does not maintain a rigid separation, although the CM still moves at an approximately constant speed ($v=19.40$ m/s at $J=2.0 \times10^{10}\,\mathrm{A/m^{2}}$), each bimeron oscillates around the CM, producing an oscillatory relative motion, see \ref{fig3}.(d). To our knowledge, this regime, in which two magnetic textures translate jointly while oscillating relative to each other, has not been previously reported under a constant applied spin-polarized current density.

\begin{figure}[ht!]
    \centering
    \includegraphics[width=0.7\linewidth]{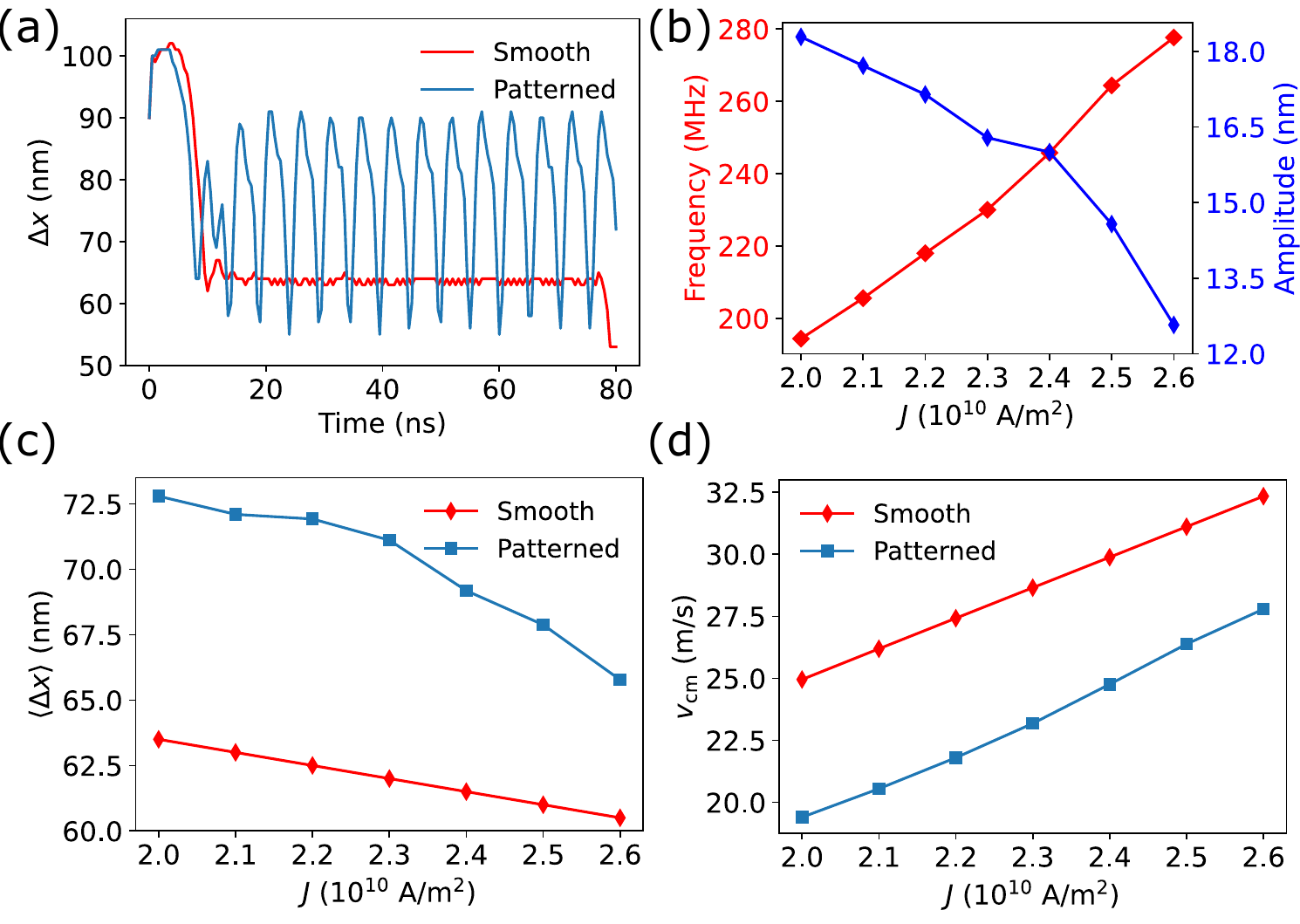}
    \caption{(a) Distance between two domain wall bimerons as a funcion of time for a current density of $J = 2.0\times 10^{10}$ A/m$^2$. (b) Frequency and amplitude of the distance between two domain wall bimerons in a patterned nanostripe as a function  of $J$. (c) Average distance between two domain wall bimerons as a function of $J$. (d) Velocity of the center of mass as a function of $J$.}
    \label{fig4}
\end{figure}

To make the oscillation reported in Figs. \ref{fig3}.(b) and \ref{fig3}.(d) explicit, we plot the instantaneous separation $\Delta x(t)$ between the two DW–bimerons in Fig. \ref{fig4}.(a). After the transient, the defect-free track (Smooth) reaches a constant separation, while the patterned track exhibits a sustained oscillation around a mean value $\langle \Delta x\rangle$. For example, we find $\langle \Delta x\rangle \approx 73~\mathrm{nm}$ at $J=2.0\times10^{10}\,\mathrm{A/m^2}$, with oscillation frequency $f\approx 194~\mathrm{MHz}$  and amplitude $A\approx 18.3~\mathrm{nm}$. Figure~\ref{fig4}.(b) shows that, as the current density increases, the oscillation frequency increases while the amplitude decreases. Figure~\ref{fig4}.(c) indicates that the time-averaged separation $\langle \Delta x\rangle$ decreases with $J$ for both smooth and patterned tracks, which supports the interpretation that the oscillation itself is triggered by the periodic defect rather than by the pair interaction alone. Consistently, Fig.~\ref{fig4}.(d) shows that the mass center velocity $v_{\mathrm{cm}}$ grows approximately linearly with $J$ in both cases, remaining lower for the patterned track. The mechanism is simple-though not entirely intuitive. The periodic defect locally reduces the velocity of the leading bimeron; when it enters a notch, $\Delta x$ decreases and the trailing bimeron (still moving faster) is drawn forward by the effective attraction between their opposite $m_x$ “charges”. As they approach, the like-signed \(m_z\) components generate a repulsive interaction that pushes them apart again. The interplay between the defect-induced velocity modulation and the combined effect of opposite $m_x$ attraction and like $m_z$ repulsion produces a bounded, self-sustained oscillation of the relative coordinate around $\langle \Delta x\rangle$ while the center of mass advances at nearly constant speed. As shown in the Supplemental Material, the oscillatory dynamics of these oscillations depend on the notch parameters, including depth, spacing, and defect width, as well as on the driving current density, indicating that both geometry and drive conditions affect the observed oscillations within the explored ranges. 

These observations can be modeled in a simple way by assuming that both bimerons are rigid particles whose positions are given by $x_1(t)$ and $x_2(t)$, and whose Lagrangian is $L=\frac1{2}\dot{x}_1^2+\frac{1}{2}\dot{x}_2^2-\frac{1}{2}\nu^2(x_1-x_2-X_O)^2$, where the interaction term is an elastic-like coupling between bimerons with strength $\nu$, and $X_O$ reflects the finite size of the bimeron. Introducing the center-of-mass and relative coordinates, $X_{cm}=\frac1{2}(x_1+x_2)$ and $R=x_1-x_2$, respectively, the Lagrangian can be rewritten as 

\begin{equation}
\label{eq:Lagrangian}
    L=\dot{X}_{cm}^2+\frac1{4}\dot{R}^2-\frac{1}{2}\nu^2(R-X_O)^2. 
\end{equation}

The Euler–Lagrange equations derived from Eq. \eqref{eq:Lagrangian} read

\begin{equation}
\label{eq:EulerLagrangeEquation}
    \ddot{X}_{cm}=0,\quad \ddot{R}=-2\nu^2(R-X_0). 
\end{equation}

The corresponding solutions are $X_{cm}(t)=X_{cm,0}+V_{cm}t$ and $R(t)=X_0+A_R\cos (\Omega t+\Phi)$, with $\Omega^2=2\nu^2$. In the original coordinates, the solutions are

\begin{eqnarray}
\label{eq:solutions}
    x_1(t)&=&X_{cm,0}+V_{cm}t+\frac1{2}X_O+\frac1{2}A_R \cos (\Omega t+\Phi), \nonumber\\
    x_2(t)&=&X_{cm,0}+V_{cm}t-\frac1{2}X_O-\frac1{2}A_R \cos (\Omega t+\Phi).
\end{eqnarray}

The inset of Fig. 3.d compares the predictions of the simplified model with the numerical results in the stationary regime, for the representative case $J=2.0\times 10^{10}$ A/m$^2$. The fitting constants obtained are $X_{cm,0}=80.52$ nm, $V_{cm}=19.41$ m/s, $A_R=18$ nm, $\Phi=-1.5287$, $X_0=73$ nm, and $\Omega=1.226$ rad/ns. These constants can be interpreted as follows: $X_{cm,0}$ is the initial position of the center of mass, $V_{cm} = v_{cm}$ corresponds to the center-of-mass velocity, $A_R$ is the amplitude of the relative oscillation between the bimerons, $\Phi$ represents the phase of the relative oscillation, and $X_0$ denotes the average separation between the bimerons, which can be identified with $\langle \Delta x \rangle$.

\begin{figure}[ht!]
    \centering
    \includegraphics[width=0.7\linewidth]{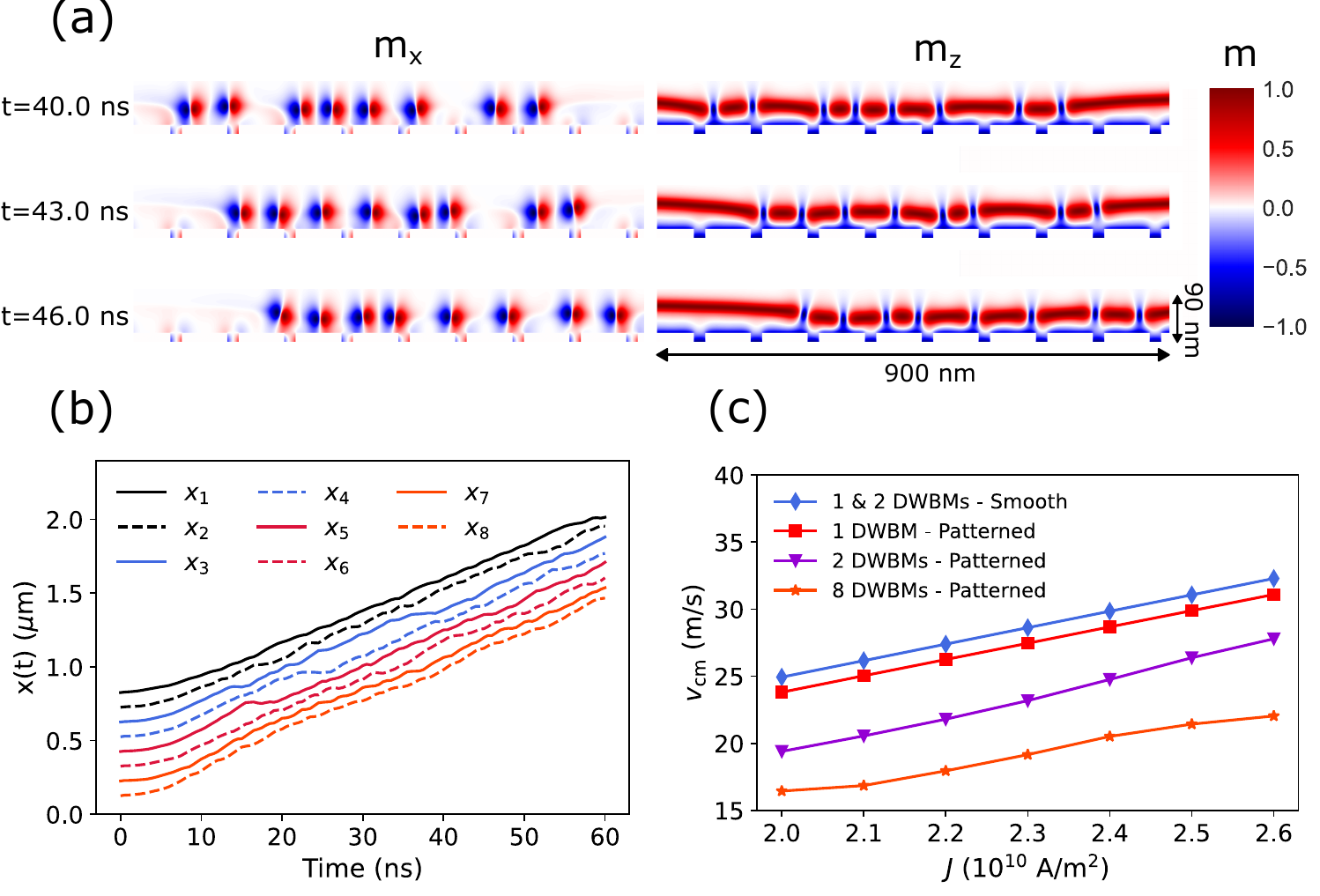}
    \caption{(a) Magnetic profiles of eight  bimerons domain wall at different time instants, stabilized in smooth and patterned nanostripes. The motion is irregular, exhibiting features of a segmented collective behavior. For further clarity about the bimeron domain wall motion see the supplementary video 3. (b) Position $x_i(t)$ of each bimeron domain wall as a function of time, where $i = 1,2, ..., 8$  labels the bimerons from left to right in the initial configuration. (c) Velocity of the center of mass of bimeron domain walls composed of different numbers of bimerons as a function of $J$, for smooth and patterned nanostripes.   }
    \label{fig5}
\end{figure}

Finally, we have all the ingredients to build and understand the magnetic
worm in a magnetic stripe. Figs.~\ref{fig5}(a)–(b) show eight bimerons in a
domain wall moving along a stripe with periodic defects. The panel b plots the
individual trajectories $x_i(t)$ of the eight bimerons. From the two bimeron case, we know that a pair translates at an almost constant mas center speed while their separation oscillates due to the combined bimeron–defect and bimeron–bimeron interactions. When the number is increased to eight, these pairwise oscillations become desynchronized and irregular, producing a segmented collective motion reminiscent of a worm. In limbless crawling organisms (e.g., worms, caterpillars, snakes, and snails), peristaltic waves may propagate either in an anterograde direction—from the posterior toward the head—or in a retrograde direction—from the head toward the tail\cite{Brackenbury1999}. Organisms that employ retrograde peristalsis, such as worms, minimize friction by elongating the contracted segments and establishing a backward-moving anchoring relative to the body\cite{Grossi2021,Silva2023}. Accordingly, the present system emulates a worm-like locomotion pattern, as the relative displacement of the anchoring moves in a rearward direction during the retrograde peristaltic wave. Figure~\ref{fig5}.(c) quantifies this behavior, i.e., the mass center speed $v_{\mathrm{cm}}$
increases with the magnitude of the applied current density $J$. However, for
fixed $J$, $v_{\mathrm{cm}}$ decreases as the number of bimerons in the domain
wall grows (compare 1, 2, and 8 bimerons in a domain wall) in a manner similar to what has been reported for skyrmion trains\cite{Xing2022}. This trend is
consistent with stronger effective drag, as more bimerons traverse the periodic
defect out of phase, their local slowdowns do not cancel, leading to alternating stretching and compression of the train and thus a lower net mass center velocity. A noteworthy point is that we could not stabilize eight bimerons in a defect-free domain wall. During the transient, one bimeron became unstable and the wall disintegrated, so this case is omitted from Fig.~\ref{fig5}.(c). In contrast, the periodic defect provides sufficient stabilization for multi-bimeron transport.

\section{Conclusions}

We studied bimerons in a domain wall driven by a spin-polarized current density  in a magnetic stripe with periodic defects. A single bimeron moves at constant speed; the defect lowers this speed but keeps an almost linear $v$–$J$ trend. With two bimerons, the center of mass moves nearly at constant speed while their separation oscillates, the frequency grows, and the amplitude shrinks as the current increases. This oscillation arises from the magnetic bimeron–defect and bimeron–bimeron interactions. With many bimerons, these oscillations lose synchrony and the train advances in a segmented. This collective behavior closely mimics the peristaltic locomotion of worms, where retrograde anchoring increase friction and enables directed motion. The mass center speed rises with current but decreases as the number of bimerons grows. The periodic defect stabilizes multi–bimeron transport (eight bimerons could not be sustained without it), enabling the speed and internal oscillations to be tuned by adjusting the drive current and the defect geometry. In practice, this enables: (i) current-controlled nano-oscillators, where the pair’s relative oscillation yields a MHz signal whose frequency is set by $J$; (ii) phase-coded racetrack logic, where bits are carried by the relative phase between the two bimerons while the center of mass transports them; and (iii) on-chip sensing of defects/pinning, where changes in oscillation frequency and mass center speed versus $J$ serve as local probes of the pinning landscape.

\begin{suppinfo}

The supplementary material includes  the dependence of the center of mass velocity and frequency of two bimerons within a domain wall as a function of the geometric parameters of the periodic defects. 
\end{suppinfo}

\begin{acknowledgement}
S.A. acknowledges funding from DICYT regular 042431AP and Cedenna CIA250002. A.S.N. acknowledges funding from Fondecyt Regular 1230515 and Cedenna CIA250002. S.C.-S. acknowledges funding from Fondecyt Regular 1251178 and Cedenna CIA250002. M.A.C. acknowledges Proyecto ANID Fondecyt Postdoctorado 3240112.  D. Galvez-Poblete acknowledges ANID-Subdirección de Capital Humano/Doctorado Nacional/2023-21230818. 

\end{acknowledgement}




\providecommand{\latin}[1]{#1}
\makeatletter
\providecommand{\doi}
  {\begingroup\let\do\@makeother\dospecials
  \catcode`\{=1 \catcode`\}=2 \doi@aux}
\providecommand{\doi@aux}[1]{\endgroup\texttt{#1}}
\makeatother
\providecommand*\mcitethebibliography{\thebibliography}
\csname @ifundefined\endcsname{endmcitethebibliography}  {\let\endmcitethebibliography\endthebibliography}{}

\end{document}